\documentclass[aps,prd,preprint,nofootinbib]{revtex4}

\usepackage{graphicx}
\usepackage{psfrag}
\usepackage{epsfig}

\begin{document}

\title{Revisiting the Real Graviton Effects at CERN LHC within the Quantum Gravity Theory with Large
Extra Dimensions}
\author{Xing-Gang Wu\footnote{email: wuxg@cqu.edu.cn} and Zhen-Yun Fang}
\address{Department of Physics, Chongqing University, Chongqing 400044,
P.R. China}

\begin{abstract}

CERN LHC provides a good experimental platform to perturbatively
probe the fundamental gravity scale up to several TeV, with the
precise value depending on the number of extra dimensions. The
leading experimental signal of graviton at LHC is from the process
$pp\to jet+{E\!\!\!\!\slash}_T$, where ${E\!\!\!\!\slash}_T$ stands
for the transverse missing energy. A detailed discussion on the
hadronic production of real graviton through hard subprocesses:
$q\bar{q}\to G+g$, $g+q\to G+q$ and $g+g\to G+g$ have been studied
within the quantum gravity theory with large extra dimensions. The
main theoretical uncertainties together with the dominant standard
model background to these processes, e.g. $q\bar{q}\to Z^0+g$ and
$g+q\to Z^0+q$ with $Z^0$ further decaying into neutrinos, have also
been discussed. It is found that only in certain jet energy region
and with certain number of extra dimensions can the quantum gravity
signal be distinguished from the background, which inversely lead to
the effective scale $M_D$ to be probed up to $(8.8\pm 0.9)$ TeV for
two extra dimensions, and $(5.9\pm 0.5)$ TeV for four extra
dimensions with sufficient integrated luminosity, e.g. $100fb^{-1}$, at CERN LHC. \\



\end{abstract}

\maketitle

\section{Introduction}

It has been purposed that standard model (SM) particles live in the
usual $3+1$-dimensional space, while gravity can propagate in a
higher-dimension space \cite{add,tev}. Such a scenario is helpful to
reduce the fundamental mass scale from the large Plank scale down to
be about $TeV$-scale and then to solve the so-called hierarchy
problem, so it arouses people's interests since its first
announcement. Numerous attempts have been carried out to find the
signal of the large extra dimension, i.e. measurements of gravity at
short distances, studies of various astrophysical and cosmological
implications of large extra dimension and collider searches for
virtual and real graviton effects
\cite{tevatron,collider,minirev,exp1,exp2,exp3,exp4,rev1}. The CERN
LHC, with its high collision energy ($14$ TeV) and high luminosity
($10^{34}cm^{-2}s^{-1}$), shall provide a better platform to study
the extra-dimension phenomenology both experimentally and
theoretically.

The leading experimental signal of real graviton at LHC is from the
hadronic process $pp\to jet+{E\!\!\!\!\slash}_T$ with
${E\!\!\!\!\slash}_T$ stands for the transverse missing energy. We
shall present a detailed discussion on the hadronic production of
graviton through the hard subprocesses: $q\bar{q}\to G+g$, $g+q\to
G+q$ and $g+g\to G+g$. By using the quantum gravity effective theory
with large extra dimensions \cite{hlz,grw}, we make a try to study
the quantum gravity effects and its dependence on the number of
extra dimension, and to study up to what energy scale can LHC probe.
Furthermore, main theoretical uncertainty for the graviton
production shall be studied, which includes the parton distribution
functions (PDFs) for the initial partons, the choice of the
factorization scale $\mu_F^2$ and the typical energy scale $Q^2$ for
the hard scattering amplitude, the number of the extra dimension for
the graviton production, and etc..

The processes $q\bar{q}\to g+Z^0$ and $qg\to q Z^0$, followed by an
invisible decay of $Z^0$, i.e. $Z^0\to\nu\bar\nu$, give an
irreducible physical background to graviton production. We refer to
these processes as the `SM background'. We will estimate the
observability of the graviton signal by comparing its hadronic
cross-section to that of `SM background'. There are other important
background sources from miss-measured jets and $W$ production with
forward leptons, however as argued in Ref.\cite{prlee} that these
backgrounds decrease sharply as the lower bound on missing $E_T$ is
increased, so we shall not considered it here.

The remainder of the paper is organized as follows. Section II is
devoted to give the main formulae for graviton production within the
framework of the quantum gravity effective theory with large extra
dimensions. Numerical results and discussions are presented in
Section III, where the uncertainties in estimates and a discussion
on the value of the effective energy scale $M_D$ shall be presented.
The last section is reserved for a summary.

\section{Calculation technology}

According to the QCD factorization formulae, the hadronic production
of graviton at the collision center of mass energy $\sqrt{S}$ can
generally be written as
\begin{eqnarray}
&&d\sigma(S,E_T,\cdots)=\nonumber \\
&&\sum_{ij}\int\int  dx_{1}dx_{2}F^{i}_{H_1,P_{1}}(x_{1},\mu^2_{F})
F^{j}_{H_2,P_{2}}(x_{2},\mu^2_{F})d\hat{\sigma}_{ij\rightarrow
jet+{E\!\!\!\!\slash}_T}(P_1,P_2,x_{1},x_{2},\mu^2_{F},Q^2,\hat{s},\cdots)\
, \label{cross}
\end{eqnarray}
where $F^{i}_{H_1,P_{1}}(x_{1},\mu^2_{F})$ and
$F^{j}_{H_2,P_{2}}(x_{2},\mu^2_{F})$ are PDFs of incoming hadrons
$H_1$ (momentum $P_1$) and $H_2$ (momentum $P_2$) for parton $i$
(with the momentum fraction $x_1$) and parton $j$ (with the momentum
fraction $x_2$) respectively. $Q^2$ is the `characteristic energy
scale of the subprocess squared'; and $\mu^2_F$ is the `energy scale
squared' where the factorization about the PDFs and the hard
subprocess is made. Usually for leading order (LO) calculation to
obtain the best results, the two scales $\mu^2_F$ and $Q^2$ are
carried out as the same, thus later on we take $\mu_F^2=Q^2$ except
one case when estimating the uncertainty from LO and the ambiguity
of the choices about $\mu_F^2$ and $Q^2$.

$d\hat{\sigma}_{ij\rightarrow jet+{E\!\!\!\!\slash}_T}$ stands for
the differential cross-section of the relevant hard subprocess, in
which $\hat{s}=x_1x_2S$ is the center of mass energy of the
subprocess and ${E\!\!\!\!\slash}_T$ stands for the missing
transverse energy. Within the framework of the quantum gravity
theory with large extra dimensions, the differential cross-section
for inclusive graviton (G) production, i.e. $ij\to Gk$ with $i$, $j$
and $k$ stands for corresponding partons, can be written as
\cite{grw}
\begin{equation}
\frac{d^2}{dt dm}\hat\sigma_{ij\to Gk} = S_{\delta -1} \frac{\bar
M_P^2} {M_D^{2+\delta}}~ m^{\delta -1}~ \frac{d\hat\sigma_m}{dt},
\label{sigf}
\end{equation}
where $d\hat\sigma_m / dt$ stands for the differential cross-section
for producing a single Kaluza-Klein graviton of mass $m$, $\bar
M_P=M_P/\sqrt{8\pi}=2.4\times10^{18}GeV$, $S_{\delta -1}$ is the
surface of a unit-radius sphere in $\delta$ dimensions, for
$\delta=2n$ and $n$ integer, $S_{\delta -1}=2\pi^n/(n-1)!$ and for
$\delta=2n+1$, $S_{\delta -1}=2\pi^n/\Pi_{k=0}^{n-1}
(k+\frac{1}{2})$. For the present $2\to2$ subprocesses, the
Mandelstam variables are defined as $s=(p_i+p_j)^2$, $t=(p_i-p_G)^2$
and $u=(p_i-p_k)^2$. Since the graviton interaction vertex is
suppressed by $1/\bar M_P$, it can be found that $\hat\sigma_m
\propto \bar M_P^{-2}$, and the factor $\bar M_P^2$ appearing from
the phase-space summation exactly cancels the Planck mass dependence
in Eq.~(\ref{sigf}). In another words, the large phase space of the
Kaluza-Klein modes, corresponding to the large volume of the
compactified space, exactly cancels the dependence on $\bar M_P$ and
gives an effective interaction suppressed only by inverse powers of
$M_D$. This is the reason why sizable contributions from the
graviton may be observed at LHC, and inversely, one can estimate to
what energy scale can LHC probe. For the differential partonic
cross-sections $d\hat\sigma_m/dt$ that produce a single Kaluza-Klein
graviton with mass $m$, we obtain
\begin{eqnarray}
\frac{d\hat\sigma_m}{dt}(q\bar q \to g G) &=&
\frac{\alpha_s(Q^2)}{36s\bar M_P^2}~F_1 (t/s,u/s), \label{csqq}\\
\frac{d\hat\sigma_m}{dt}(qg \to q G) &=&
\frac{\alpha_s(Q^2)}{96s\bar M_P^2}~F_2 (t/s,u/s), \label{csqg}\\
\frac{d\hat\sigma_m}{dt}(gg \to g G) &=&
\frac{3\alpha_s(Q^2)}{16s\bar M_P^2}~F_3 (t/s,u/s), \label{csgg}
\end{eqnarray}
where $Q^2$ stands for the characteristic energy scale of the hard
scattering amplitude, $q$ stands for the light quark $u$, $d$ and
$s$ respectively, and the functions $F_{1,2,3}$ take the following
form
\begin{eqnarray}
F_1(x,y)&=& \frac{1}{xy}(4xy + z)(1 - 2xy + z^2), \\
F_2(x,y)&=& \frac{1}{xz}(2x-z^2 - y^2 )( z + z^2 + x( 4 + z)) , \\
F_3(x,y)&=&\frac{1}{xz} \left[x^2 y^2 + 2 x y(z^2-z+1) +( 1 + z +
z^2)^2 \right].
\end{eqnarray}
The relation $z=1+x+y$ is implicitly adopted that can be deduced
from the fact of $s+t+u=m^2$. It can be easily find that
$F_{1,3}(x,y)=F_{1,3}(y,x)$ as is the requirement of the invariance
under exchange of the Mandelstam variables $t$ and $u$.

For the background subprocesses $ij\to Z^0 k$, we obtain
\begin{eqnarray}
\frac{d\hat\sigma}{dt}(q\bar q \to g Z^0) &=&
\frac{\pi\alpha(Q^2)\alpha_s(Q^2)(L_q^2+R_q^2)}{9x_W(1-x_W)s^2tu}
(m_{Z^0}^4+s^2-2tu), \label{cszqq}\\
\frac{d\hat\sigma}{dt}(qg \to q Z^0) &=&
\frac{\pi\alpha(Q^2)\alpha_s(Q^2)(L_q^2+R_q^2)}
{18x_W(1-x_W)m_{Z^0}^2s^3t} [6sm_{Z^0}^4-2s(s+t)m_{Z^0}^2+t^3],
\label{cszqg}
\end{eqnarray}
where $x_W=\sin^2\theta_W$, $R_q=-2e_q x_W$ and $L_q=1/2-2e_q x_W$
with $e_q$ stands for the electric charge of $q$-quark.

\section{Numerical results and discussions}

In the following, we shall first discuss the uncertainties in
estimating the graviton/$Z^0$ production. For the LO estimation, we
shall concentrate our attention on the main uncertainties, which
include the PDFs for the light quarks, the choice of the
factorization scale $\mu_F$ and the typical energy scale $Q^2$ for
the hard scattering amplitude and the number of the extra dimension
for the graviton production.

The number of extra dimension ($\delta$) can either be too small or
too large. We are interested in the case in which $\delta$ is not
too large (say $\delta \lesssim 6$), under such condition the mass
splitting $\Delta m$ is so small that the sum over the different
Kaluza-Klein states can be replaced by a continuous integration, and
then the enormous number of accessible Kaluza-Klein modes can
rightly compensate the $1/\bar{M}_P^2$ factor in the scattering
amplitude. While an even larger number of extra dimensions shall
lead to the mass splitting $\Delta m$ become comparable with the
experimental energy resolution, hence only a smaller number of
Kaluza-Klein modes can be produced, and then the total cross-section
is negligible due to the unavoidable $1/\bar{M}_P^2$ suppression. On
the other hand, it is argued that $\delta$ can not be too small as
is required by the latest torsion-balance experiment \cite{torsion}.
Constraints from cosmology also lead to non-trivial bounds on extra
dimensions \cite{cosmo1,cosmo2} \footnote{By using a hyperbolic
curved other than the flat extra dimensions \cite{hyper}, it has
been argued that those cosmology constraints can be naturally
satisfied \cite{mel}.}. So, we take $\delta \in [2, 6]$ to do study
its effect to the hadronic production.

As is shown in Eq.(\ref{cross}), PDFs $F^{i}_{H_1,
P_{1}}(x_{1},\mu^2_{F})$ and $F^{j}_{H_2, P_{2}}(x_{2},\mu^2_{F})$
generate certain uncertainties in the estimation. PDFs are of
non-perturbative nature, and in Eq.(\ref{cross}) they have been
factorized out at the energy scale $\mu_F^2$ with the help of pQCD
factorization theorem. The PDFs can be obtained only through global
fitting of the experimental data and evolute them to the requested
characteristic scale in standard way of pQCD, so there are several
groups, e.g. CTEQ \cite{6lcteq}, GRV \cite{98lgrv} and MRS
\cite{2001lmrst} etc, who devote themselves to offer accurate PDFs
to the world, and to keep PDFs updated with the newly available
relevant experimental data. Thus in literature, different versions
of PDFs (including different issues by the same group) are used in
the estimates of the hadronic production. To be self-consistent with
the LO pQCD calculation, we shall adopt the two LO PDFs: CTEQ6L
\cite{6lcteq} and MRST2001L\cite{2001lmrst} as typical examples for
PDFs. The versions of the gluon distributions ended with `L' are
accurate up-to the leading logarithm order (LLO), i.e., their QCD
evolution effects with $\alpha_s$ running are included, so for the
production to show the uncertainties correctly up to LO accuracy, it
is necessary for the PDFs, the hard subprocess and the QCD `coupling
constant' $\alpha_s$ `run' to the energy scale $Q^2$ properly. When
computing the production and taking the PDFs from one version of the
three groups, the running $\alpha_s$ should also be taken from the
same group.

As for the leading order estimation, how to choose the energy scale
$Q^2$ is a very tricky problem. If $Q^2$ is chosen properly, the
results may be quite accurate. From experience, for a hard
subprocess with two-body final state, generally the choice of
$Q^2=\hat{s}/4$ can lead to an accurate LO result. To see the
uncertainty caused by different choices of $Q^2$, we take two
typical types for $Q^2$, i.e. Type $A$: $Q^2=\hat{s}/4$ with
$\hat{s}$ the squared center of mass energy of the subprocess; Type
$B$: $Q^2=M_t^2\equiv p_{t}^2+m^2$, the squared transverse mass of
the graviton/$Z^0$ respectively. And further more, To see the
uncertainties from $Q^2$ choice, instead of variation on the choices
with $Q^2=\mu_F^2$, the authors in literature, such as
Ref.\cite{psi}, also try $Q^2\neq \mu_F^2$ and see the uncertainty.
Here following them, we also calculate the distributions with
$Q^2\neq \mu_F^2$. More explicitly, as suggested in
Refs.\cite{psi,cww}, we take $\mu^2_F\in [M_{t}^2/4, 4M_{t}^2]$ and
$Q^2\in [M_{t}^2/4, 4M_{t}^2]$ simultaneously to do the discussion.

Other parameters like $m_{Z^0}$, $x_W$ and the fraction of $Z^0$
decaying to invisible particles affect the production slightly, so
we directly take them to be their center values as adopted in the
literature, e.g. $m_{Z^0}=91.187$GeV, $x_w$ to be $0.2311$ and the
fraction of $Z^0$ decaying to invisible particles like
(anti)neutrinos $\nu (\bar\nu)$ to be the center value of
$(20.00\pm0.06)\%$ \cite{pdg}.

Another important thing we need to be careful is the regulation of
the cross section to avoid the non-perturbative regime, i.e. to deal
with the signatures arising from collisions with parton
center-of-mass energies of order $M_D$ or above properly. At
$\sqrt{\hat{s}}\gg M_D$, parton collisions are expected to produce
classical black holes \cite{Bourilkov}. As has been pointed out that
the discrepancy between the two cases with or without the cut
$\hat{s}\leq M^2_D$ shall be increased with the increment of
$E^{min}_{T,jet}$ and decreased with the increment of $M_D$
\cite{grw}, therefore by taking proper $E_{T,jet}$ cut in
experimental analysis, one can select different ranges of $M_D$ to
probe perturbatively. And numerically, we can find that under the
present conditions, with or without the cut $\hat{s}\leq M^2_D$
shall affect our final conclusions slightly, so we shall adopt the
cut $\hat{s}\leq M^2_D$ in all the following discussions.

\subsection{The uncertainties in estimates}

\begin{figure}
\centering
\includegraphics[width=0.55\textwidth]{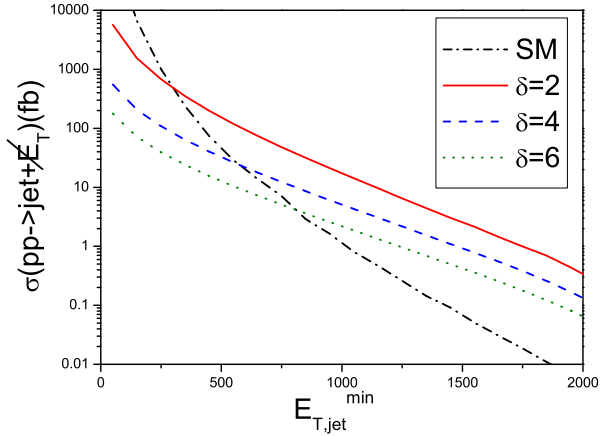}
\caption{Total jet+nothing cross-section at LHC integrated for the
requirement that $E_{T,jet}>E^{min}_{T,jet}$ with an acceptance cut
on the jet rapidity $|\eta|\leq 3$. The dash-dotted line is SM
background, the solid, dashed and dotted lines are for $\delta=2$,
$4$ and $6$ respectively. } \label{dimension}
\end{figure}

Firstly, we show the uncertainty caused by the different choice of
extra dimensions $\delta$. For the purpose, we fix PDF to be CTEQ6L,
$M_D=5TeV$, $Q^2=\hat{s}/4$ and $\mu_F^2=Q^2$. We show the signal
and the background rates for the transverse jet energy larger than
$E^{min}_{T,jet}$ in FIG.\ref{dimension}, with an acceptance cut on
the jet rapidity $|\eta|\leq 3$. It is shown that the signal rates
decrease with the increment of $\delta$. The SM background is bigger
than the graviton signal in the lower transverse jet energy region
but it drops down much quickly than the signal, and one may
distinguish the signal from the background in large transverse jet
energy region. In another words, the large transverse jet energy
region shall provide an effective platform to distinguish the signal
and the background.

\begin{figure}
\centering
\includegraphics[width=0.55\textwidth]{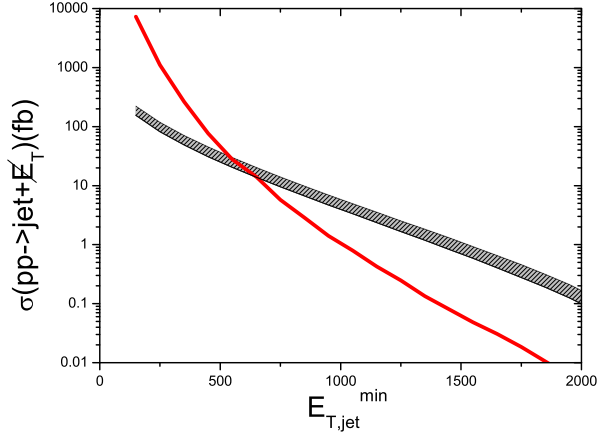}
\caption{Total jet+nothing cross-section at LHC integrated for the
requirement that $E_{T,jet}>E^{min}_{T,jet}$ with an acceptance cut
on the jet rapidity $|\eta|\leq 3$. The thinner and thicker shaded
bands stand for the SM background and the signal respectively, with
the upper edge for $\mu_F^2=M_t^2/4$ and $Q^2=M_t^2/4$, and lower
edge for $\mu_F^2=4M_t^2$ and $Q^2=4M_t^2$. } \label{scale}
\end{figure}

Secondly, we show the uncertainty from different choices of $Q^2$
and $\mu_F^2$. For such purpose, we fix PDF to be CTEQ6L,
$M_D=5TeV$, $\delta=4$. It is found that by taking two choices of
$Q^2$ (type A and type B) under the case of $\mu_F^2=Q^2$, the
uncertainties for both the background and the signal are small, i.e.
the differences between these two types of $Q^2$ are less than 10\%
for both the background and the signal. Further more, we show the
case of $\mu_F^2\neq Q^2$ by shaded band in FIG.\ref{scale}, with an
acceptance cut on the jet rapidity $|\eta|\leq 3$, where $\mu^2_F\in
[M_{t}^2/4, 4M_{t}^2]$ and $Q^2\in [M_{t}^2/4, 4M_{t}^2]$. It is
found the largest value is obtained when $\mu_F^2=M_t^2/4$ and
$Q^2=M_t^2/4$, and the lowest value is obtained when
$\mu_F^2=4M_t^2$ and $Q^2=4M_t^2$. And from FIG.\ref{scale} one may
also observe that while the uncertainty for the background changes
to be around $20\%$ (as shown by the thinner shaded band), the
uncertainty for the signal can be changed up to 50\% (as shown by
the thicker shaded band).

\begin{figure}
\centering
\includegraphics[width=0.55\textwidth]{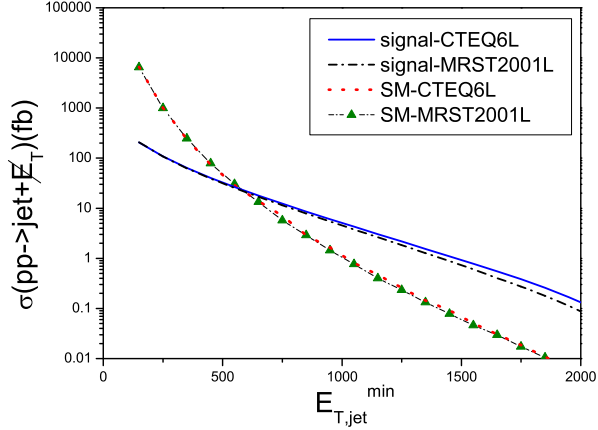}
\caption{Total jet+nothing cross-section at LHC integrated for the
requirement that $E_{T,jet}>E^{min}_{T,jet}$ with an acceptance cut
on the jet rapidity $|\eta|\leq 3$. Two typical LO PDFs, e.g.
CTEQ6L, and MRST2001L are adopted. } \label{pdf}
\end{figure}

\begin{table}
\begin{center}
\caption{Total jet+nothing cross-section $\sigma(pp\to
jet+\slash\!\!\! E_{T,jet})$ (in unit: fb) versus PDFs at LHC
integrated for the requirement that $E_{T,jet}>E^{min}_{T,jet}$ with
an acceptance cut on the jet rapidity $|\eta|\leq 3$. Two LO PDFs:
CETQ6L and MRST2001L, and one NLO PDF: CTEQ6M, are adopted.} \vskip
0.6cm
\begin{tabular}{|c||c|c|c||c|c|c||}
\hline - & \multicolumn{3}{|c||}{~~~Signal with $E_{T,jet}>E^{min}_{T,jet}$~~~~}
& \multicolumn{3}{|c||}{~~~SM background with $E_{T,jet}>E^{min}_{T,jet}$~~~~}\\
\hline ~~~PDFs~~~ & ~~$0.50$ TeV~~ & ~~$1.0$ TeV~~ & ~~$1.5$ TeV~~ &
~~ $0.50$ TeV~~ & ~~$1.0$ TeV~~ & ~~$1.5$ TeV~~\\
\hline\hline CTEQ6L & 32. & 5.1 & 0.92 & 46. & 1.1 & 0.070\\
\hline MRST2001L & 31. & 4.5 & 0.73 & 47. & 1.0 & 0.060\\
\hline\hline CTEQ6M & 40. & 6.3 & 1.1 & 57. & 1.1 & 0.074\\
\hline\hline
\end{tabular}
\label{tabpdf}
\end{center}
\end{table}

Thirdly, we show the uncertainty from different choices of PDFs by
fixing $M_D=5TeV$, $\delta=4$ and $\mu_F^2=Q^2=\hat{s}/4$. We show
the results for two LO PDFs in FIG.\ref{pdf} with an acceptance cut
on the jet rapidity $|\eta|\leq 3$, i.e. CTEQ6L and MRST2001L. It is
found that the results of CTEQ6L and MRST200L are close to each
other, i.e the difference is less than $20\%$. More explicitly, we
show the total jet+nothing cross-section $\sigma(pp\to
jet+\slash\!\!\! E_{T,jet})$ versus PDFs at LHC in TAB.\ref{tabpdf}.
TAB.\ref{tabpdf} shows more explicitly that even though the SM
background is bigger than the graviton signal in the lower
transverse jet energy region but it drops down quickly, and at
$E^{min}_{T,jet}\sim 1TeV$ it is less than $25\%$ of the signal. At
the present the next-to-leading order (NLO) results are not
available \footnote{Such a NLO calculation is in
progress\cite{wu}.}, and to have a rough idea on how NLO calculation
will affect the present results, we take the `miss matched' NLO PDF
as CTEQ6M \cite{6lcteq} to do the calculation. And it is found that
the results from the `miss marched' CTEQ6M shall be bigger than that
of CTEQ6L by about $25\%$. So a full NLO estimation shall be helpful
to improve our present estimations.

\begin{figure}
\centering
\includegraphics[width=0.55\textwidth]{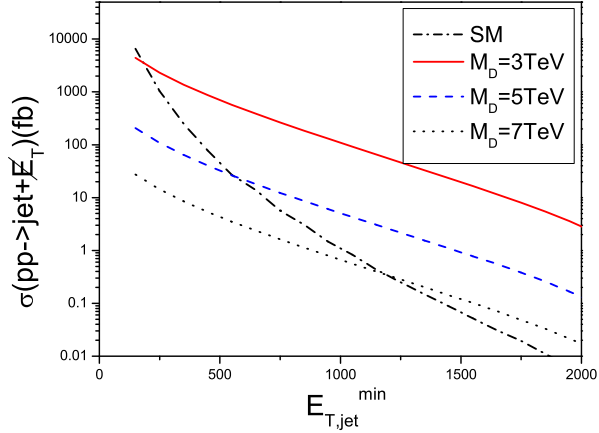}
\caption{Total jet+nothing cross-section at LHC integrated for the
requirement that $E_{T,jet}>E^{min}_{T,jet}$ with an acceptance cut
on the jet rapidity $|\eta|\leq 3$. Three typical choices for
$M_D=3$GeV, 5GeV and 7GeV are adopted. } \label{md}
\end{figure}

Finally, we show the uncertainty from different choices of $M_D$ by
fixing PDF to be CTEQ6L, $\delta=4$, $Q^2=\hat{s}/4$ and
$\mu_F^2=Q^2$. We show the signal and the background rates for the
transverse jet energy larger than $E^{min}_{T,jet}$ in FIG.\ref{md}
with an acceptance cut on the jet rapidity $|\eta|\leq 3$. It is
shown that the rate decreases with the increment of $M_D$ under the
same condition $E_{T,jet}>E^{min}_{T,jet}$.

\subsection{To what energy scale can LHC probe ?}

\begin{table}
\begin{center}
\caption{Corresponding $M_D$ sensitivity ranges versus $\delta$ with
the high integrated luminosity ${\cal L}=100fb^{-1}$. $M_D
(3\sigma)$ stands for the $3\sigma$ exclusion limit and $M_D
(5\sigma)$ stands for $5\sigma$ observable limit, where the center
values are obtained by taking CTEQ6L, $Q^2=\hat{s}/4$ and
$\mu_F^2=Q^2$ and the errors are caused by the above mentioned main
uncertainty sources. } \vskip 0.6cm
\begin{tabular}{|c||c|c||}
\hline ~~~$\delta$~~~ & ~~~$M_D (3\sigma-exclusion)$~~~ & ~~~$M_D (5\sigma-observable)$~~~ \\
\hline\hline 2 & ~~~$9.2\pm 0.7$ TeV~~~ & ~~~$8.8\pm 0.9$ TeV~~~ \\
\hline 4 & $6.1\pm 0.5$ TeV & $5.9\pm 0.5$ TeV  \\
\hline 6 & $5.3\pm 0.3$ TeV & $5.1\pm 0.3$ TeV  \\
\hline\hline
\end{tabular}
\label{tabscale}
\end{center}
\end{table}

It is found that the SM background changes slightly within the
reasonable regions of the above mentioned uncertain sources, so it
can be treated as a basis to decide to what energy scale can LHC
probe. Since the large extra dimensions can be probed only when the
deviation of the cross section within the framework of large extra
dimension model from the SM background is large enough, we adopt the
$5\sigma$ large extra dimension effect observable limit and
$3\sigma$ exclusion limit as suggested in the literature \cite{ma}
to extract the constraint of fundamental energy scale $M_D$, i.e.
\begin{eqnarray}
\Delta\sigma &=& \sigma_{LED}-\sigma_{Bgd}\geq
\frac{5\sqrt{\sigma_{LED} {\cal L}}}{\cal L} \label{cr1}\\
\Delta\sigma &=& \sigma_{LED}-\sigma_{Bgd}\leq
\frac{3\sqrt{\sigma_{LED} {\cal L}}}{\cal L}. \label{cr2}
\end{eqnarray}
The corresponding $M_D$ sensitivity ranges versus $\delta$ are shown
in TAB.\ref{tabscale} with high integrated luminosity ${\cal
L}=100fb^{-1}$, where the cuts $|\eta|\leq 3$ and $E_{T,jet}>1.0$
TeV are adopted. It has found that with integrated luminosity ${\cal
L}\sim$ several $fb^{-1}$, the main uncertainty comes from the
instrumental background \cite{exp1,exp2,exp3,exp4}, which includes
both the systematic and the statistical errors. So we have taken a
higher integrated luminosity ${\cal L}=100fb^{-1}$ to do our
calculation such that the systematic error is dominant. And in doing
the calculation, we require that $\sigma_{LED}>\sigma_{Bkgd}$, since
it has been found that the present adopted effective gravity theory
is mostly reliable in this region as shown in Ref.\cite{grw}. It is
found that the effective scale $M_D$ can be probed up to $(8.8\pm
0.9)$ TeV for $\delta=2$, $(5.9\pm 0.5)$ TeV for $\delta=4$ and
$(5.1\pm 0.3)$ TeV for $\delta=6$, where the center values are
obtained by taking CTEQ6L, $Q^2=\hat{s}/4$ and $\mu_F^2=Q^2$ and the
errors are caused by the above mentioned main uncertainty sources
that varies within their reasonable regions accordingly.

Further more, one may observe that the center values for $M_D$
decreases with the increment of $\delta$. Our present results for
$M_D$ versus $\delta$ with $5\sigma$ observable limit as shown in
TAB.\ref{tabscale} are consistent with the values of maximum $M_D$
determined in Ref.\cite{grw} (TAB.3 there for the same integrated
luminosity ${\cal L}=100fb^{-1}$) within reasonable uncertainties,
with the center value of our present one slightly bigger than that
of Ref.\cite{grw}, which is mainly caused by the fact that different
criterion was adopted in Ref.\cite{grw}, i.e. a fixed systematic
error that is about $10\%$ is adopted to do the discussion.

It has been argued that \cite{grw} if the discrepancy for the
results with or without the cut $\hat{s}\leq M^2_D$ becomes larger,
then the ultraviolet contributions become important, and then our
present estimation may be not under control. Numerically, we find
that such discrepancy is small for $\delta \leq 4$ (e.g. it is less
than $1\%$ for $\delta=2$ and $10\%$ for $\delta=4$) by taking the
$M_D$ values listed in TAB.\ref{tabscale}, while for even larger
$\delta$ such discrepancy becomes quite large, e.g. for $\delta=6$
such discrepancy is up to $100\%$ by taking $M_D=5.1$ TeV. This
shows that our present adopted effective theory may not be fully
applicable for such a large extra dimension. One may hope to
decrease such discrepancy by lowering the value of $E^{min}_{T,jet}$
since a smaller $E^{min}_{T,jet}$ leads to a smaller discrepancy,
however by doing this, the probed $M_D$ sensitivity range shall only
be slightly lowered due to the fact that $E^{min}_{T,jet}$ can not
be set too small otherwise it will be more difficult to distinguish
the signal from the background, as has been shown in the last
subsections that the background shall increase much more quickly
than the signal with a decreasing $E^{min}_{T,jet}$.

\begin{table}
\begin{center}
\caption{Maximum $M_D$ sensitivity versus $\delta$ with the low
integrated luminosity ${\cal L}=10fb^{-1}$. The center values are
obtained by taking CTEQ6L, $Q^2=\hat{s}/4$ and $\mu_F^2=Q^2$ and the
errors are caused by the above mentioned main uncertainty sources. }
\vskip 0.6cm
\begin{tabular}{|c||c|c|c||}
\hline ~~~$\delta$~~~ & ~~~$2$~~~ & ~~~$4$~~~ & ~~~$6$~~~ \\
\hline\hline Max $M_D$ & ~~~$8.2\pm 0.5$ TeV~~~ & ~~~$5.7\pm 0.3$ TeV~~~& ~~~$4.9\pm 0.2$ TeV~~~ \\
\hline
\end{tabular}
\label{tab3}
\end{center}
\end{table}

It may be also interesting to make a discussion on the maximum $M_D$
sensitivity with smaller integrated luminosity, e.g. $10fb^{-1}$.
Under such case both symmetric and statistical errors are
comparable, and now the criterions (\ref{cr1},\ref{cr2}) are not
applicable, so we adopt the criterion suggested by Ref.\cite{grw} to
do the discussion, i.e. we add the two errors in quadrature and
require
\begin{equation}
\sigma_{LED}>\sqrt{2}\frac{5\sqrt{\sigma_{Bgd} {\cal L}}}{{\cal L}}.
\end{equation}
The corresponding $M_D$ sensitivity ranges versus $\delta$ are shown
in TAB.\ref{tab3} with lower integrated luminosity ${\cal
L}=10fb^{-1}$, where the cuts $|\eta|\leq 3$ and $E_{T,jet}>1.0$ TeV
are adopted and the errors are caused by the above mentioned main
uncertainty sources that varies within their reasonable regions
accordingly.

\section{Summary}

It is found that with sufficient luminosity at LHC the fundamental
gravity scale can be probed up to several TeV, with the precise
value depending on the number of extra dimensions. In the present
paper, we have presented a detailed discussion on the leading
experimental signal of real graviton at LHC based on the process
$pp\to jet+\slash\!\!\!\! E_T$ with the help of the quantum gravity
theory with large extra dimensions. The main standard model
background to these processes together with their uncertainties have
also been discussed. It is found that in higher transverse jet
energy region, e.g. $E_{T,jet}>1.0$ TeV, and with certain number of
extra dimensions, the quantum gravity signal can be distinguished
from the background.

\hspace{0.6cm}

\noindent{\bf Acknowledgements: } We would like to thank Prof. F.Y.
Li for helpful discussions. This work was supported in part by
Natural Science Foundation Project of CQ CSTC under grant number
2008BB0298 and Natural Science Foundation of China under grant
number 10805082, and by the grant from the Chinese Academy of
Engineering Physics under the grant numbers: 2008T0401 and
2008T0402. \\

\end{document}